\documentclass[twocolumn,runningheads,final]{svjour2}
\smartqed  
\usepackage{graphicx}

\newcommand{\pd}{\partial}

\newcommand{\beq}{\begin{equation}}
\newcommand{\eeq}{\end{equation}}
\newcommand{\beqn}{\begin{eqnarray}}
\newcommand{\eeqn}{\end{eqnarray}}
\newcommand{\lppr}{\stackrel{<}{\scriptstyle \sim}}
\newcommand{\gppr}{\stackrel{>}{\scriptstyle \sim}}

\journalname{Astrophysics and Space Science}
\begin{document}

\title{Fermi acceleration in astrophysical jets}

\titlerunning{Fermi acceleration}

\author{Frank M. Rieger  \and
       Valent\'{\i} Bosch-Ramon \and 
       Peter Duffy
}

\authorrunning{Rieger, Bosch-Ramon and Duffy} 

\institute{F.M. Rieger \at
              UCD School of Mathematical Sciences \\
              University College Dublin, Belfield, Dublin 4, Ireland\\
              \email{frank.rieger@ucd.ie}           
           \and
           V. Bosch-Ramon \at
              Departament d'Astronomia i Meteorologia\\ 
              Universitat de Barcelona, Av. Diagonal 647\\ 
              08028 Barcelona, Spain
           \and 
           P. Duffy \at
              UCD School of Mathematical Sciences \\
              University College Dublin, Belfield, Dublin 4, Ireland
}

\date{Received: date / Accepted: date}

\maketitle

\begin{abstract}
We consider the acceleration of energetic particles by Fermi 
processes (i.e., diffusive shock acceleration, second order Fermi
acceleration, and gradual shear acceleration) in relativistic 
astrophysical jets, with particular attention given to recent 
progress in the field of viscous shear acceleration. We analyze 
the associated acceleration timescales and the resulting particle 
distributions, and discuss the relevance of these processes for the 
acceleration of charged particles in the jets of AGN, GRBs and 
microquasars, showing that multi-component powerlaw-type particle 
distributions are likely to occur.

\end{abstract}

\section{Introduction}
Ever since the earliest detections of non-thermal emission from 
jet-type astrophysical sources astrophysicists have conjectured 
upon its origin. Today, it is widely believed that Fermi processes, 
where particle acceleration occurs as a consequence of multiple 
scattering of energetic particles off magnetic turbulence with 
a small energy change in each event, are responsible for the 
production of the non-thermal powerlaw particle distributions as 
required by the observed synchrotron and inverse Compton emission 
properties of these jets.
First-order Fermi acceleration at strong nonrelativistic shocks,
observationally well established to take place in the shells of 
supernova remnants (e.g., Aharonian et al.,~2004), can, for example, 
naturally account for the commonly required powerlaw particle 
spectra $N(\gamma) \propto \gamma^{-s}$ with spectral indices $s 
\simeq 2$ and is also a sufficiently fast and efficient mechanism.
On observational grounds such an interpretation is strongly supported 
by (i) the fact that the knotty features detected in extragalactic 
jets can be directly identified with sites of strong shock formation, 
and (ii) by the multiple detection of characteristic variability 
patterns (e.g., spectral index hysteresis) associated with efficient 
first-order Fermi acceleration in AGN-type jets (Kirk et al., 1998).
Recent high-resolution studies of extragalactic jets, however,
indicate that first-order Fermi acceleration alone, localized
by its very nature, cannot satisfactorily account for the
detection of extended high-energy emission. In the case of the
quasar 3C~273, for example, the optical spectral index is found
to vary only smoothly along the (large-scale) jet with no signs
of strong synchrotron cooling at any location in the jet, e.g.,
between knots, contrary to expectations from shock acceleration
scenarios (Jester et al., 2001; Jester et al., 2005), thus 
suggesting the need of a continuous stochastic re-acceleration 
mechanism operating all along the jet.
Shear and/or second-order Fermi particle acceleration, although 
possibly swamped by first-order Fermi processes in the vicinity 
of a flow discontinuity, appear to be the most natural candidates 
that may account for these observations.\\
Here we analyse some of the essential properties of Fermi particle 
acceleration processes and discuss their relevance for different 
astrophysical jet sources. We wish to note that we are focusing on 
mechanisms operating {\it within} a jet and not at its working 
surface (hot spot).

\section{Fermi acceleration processes}\label{fermi}
Fermi particle acceleration (Fermi, 1949) is essentially
based on the fact the energetic particles (velocity $v \sim c$)
can gain energy by elastically scattering off magnetic turbulence
structures or irregularities moving with some characteristic
velocity $\vec{u}$. Following a simple microscopic treatment and
assuming energy to be conserved in the comoving scattering frame,
the energy change of a particle due to collision is simply given
by
\beq\label{gain}
\Delta \epsilon:=
\epsilon_2 -\epsilon_1 = 2 \Gamma^2 (\epsilon_1 u^2/c^2
                         -\vec{p}_1 \cdot \vec{u})\,,
\eeq where $\Gamma=(1-u^2/c^2)^{-1/2}$ is the Lorentz factor,
$\vec{p}=E \vec{v}/c^2$ the particle momentum and the indices 1
and 2 denote particle properties before and after scattering.
A particle thus gains or loses energy depending on whether it
suffers head-on/approaching ($\vec{p_1} \cdot \vec{u} < 1$)
or following/overtaking ($\vec{p_1} \cdot \vec{u} > 1$) collisions.
Based on these considerations the following cases may be
distinguished:

\subsection{Shock or first-order Fermi acceleration}
Suppose that a strong (nonrelativistic) shock wave propagates
through the plasma. Then in the frame of the shock the conservation
relations imply that the upstream velocity (ahead of the shock) is
much higher than the downstream velocity (behind the shock), i.e.,
$u_u/u_d = (\gamma_h+1)/(\gamma_h-1)$, with $\gamma_h >1$ denoting 
the ratio of specific heats, so that the two regions may be regarded
as two converging flows. Hence, in the upstream [downstream] rest
frame the plasma from the other side of the shock (downstream
[upstream]) is always approaching with velocity $u=u_u-u_d$, so
that to first order there are only head-on collisions for
particles crossing the shock front. The acceleration process,
although stochastic, thus always leads to a gain in energy, so 
that for magnetic turbulence structures virtually comoving with 
the plasma flow, the energy gain [eq.~(\ref{gain})] becomes first 
order in $u/c$, i.e., 
\beq
 \frac{\Delta \epsilon}{\epsilon_1} \propto \frac{u}{c}\,.
\eeq
The acceleration timescale $t_{\rm acc}$ for diffusive shock
acceleration depends on both, the upstream and downstream residence
times (Drury, 1983). In general, for a useful order of magnitude
estimate $t_{\rm acc} \sim 3 \kappa/u_s^2$ (Kirk and Dendy, 2001),
with $u_s$ the shock speed measured in the upstream frame and
$\kappa$ the spatial diffusion coefficient. Particularly, for 
quasi-parallel shocks with $\kappa = \kappa_d \simeq \kappa_u$ one 
finds $t_{\rm acc} \simeq 20\,\kappa/u_s^2$ (e.g., Protheroe and 
Clay, 2004), so that in the quasi-linear limit $|\delta B| \lppr B$, 
where $\kappa \lppr \kappa_B \simeq r_g\,c/3$, 
\beq\label{t_shock}
t_{\rm acc} \gppr 6 \frac{\gamma\,m\,c}{e\,B} 
                  \left(\frac{c}{u_s}\right)^2\,.
\eeq In theory, much faster acceleration may be achieved for 
quasi-perpendicular shocks, where -- assuming quasi-linear 
approximation ($|\delta B| \ll B$) to hold -- the (perpendicular) 
diffusion coefficient $\kappa$ can be significantly smaller than 
the above quoted Bohm limit (Jokipii, 1987). However, for realistic
astrophysical applications such a situation seems at least 
questionable given recent numerical results which show that cosmic 
ray streaming at a shock front can lead to strong self-generated 
turbulence beyond the quasi-linear regime (Lucek and Bell, 2000).\\  
Fermi acceleration at (unmodified) nonrelativistic shocks is known
to produce powerlaw particle spectra $N(\gamma) \propto \gamma^{-s}$, 
which are essentially independent of the microphysics involved and 
only dependent on the shock compression ratio $\rho=u_u/u_d$ (where 
$1 < \rho \leq 4$), i.e., 
\beq
  s=\frac{(\rho+2)}{(\rho-1)}\,,
\eeq so that for strong shocks ($\rho=4$ in the test particle limit) 
the famous $s=2$ result is obtained (Drury, 1983; Blandford and Eichler, 
1987). Note that incorporation of non-linear effects (e.g., strong 
shock modification) usually suggests values $s < 2$ at high energies
(Berezhko and Ellison, 1999). On the other hand, incorporation of 
anomalous (non-diffusive) transport properties associated with the 
wandering of magnetic field lines, may efficiently reduce cross-field 
propagation and thus allow values up to $s=2.5$ (Kirk et al., 1996).\\ 
To undergo efficient first-order Fermi acceleration at non-relativistic 
shocks electrons already have to be preaccelerated up to seed Lorentz 
factors $\gamma_e > m_p/m_e\,(V_A/c)$ (ion cyclotron resonance condition). 
Recent simulations suggest that this "problem of injection" may possibly 
be resolved by electrostatic wave (ESW) surfing  -- when ESWs, excited by 
streaming ion beams, saturate by trapping electrons, thus transporting
them across the magnetic field --  and/or acceleration due to ESW collapse 
(McClements et al., 2001; Dieckmann et al., 2004).

\subsection{Second-order Fermi acceleration}
Suppose that the scattering centres have a non-negligible random
velocity component. In the absence of dominant shock effects (see
above), energetic particles will thus experience both head-on and
overtaking collisions, i.e., lose and gain energy. However, as the
rate of collisions is proportional to $|\vec{v_1}-\vec{u}|/v_1
\simeq (1-\vec{v}_1 \vec{u}/v_1^2)$, there is a higher probability
for head-on compared to overtaking collisions, which gives an average
energy gain per collision that is second order in $u/c$, i.e.,
\beq
 \frac{<\Delta\epsilon>}{\epsilon_1} \propto 
                         \left(\frac{u}{c}\right)^2 
\eeq when averaged over all momentum directions. Second-order Fermi 
acceleration thus represents a classical example of a stochastic 
acceleration process due to many small, nonsystematic energy changes. 
As such it can be described by a diffusion equation in momentum space 
(Skilling, 1975; Melrose, 1980), i.e., the isotropic phase space 
distribution (averaged over all momentum directions) evolves according 
to
\beq\label{diffusion}
 \frac{\pd f(p)}{\pd t} = \frac{1}{p^2}\frac{\pd}{\pd p}
                 \left(p^2 D_{p}(p) \frac{\pd f(p)}{\pd p}\right)\,,
\eeq with $D_{p}(p) \propto <(\Delta p)^2>$ the diffusion coefficient
in momentum space. A statistical treatment taking the small anisotropy
of the particle distribution in the laboratory frame into account
allows proper calculation of the Fokker-Planck coefficients (Duffy
and Blundell, 2005) and for $v \sim c$ gives $D_p =(u/c)^2 p^2/(3 \tau)
\propto p^2/\tau$, where $\tau \simeq 1/(n\,\sigma\,c)$ is a mean
scattering time and $n$ is the number density of scatterers. For
the scattering off forward and reverse propagating Alfven waves a
similar expression can be derived (Skilling, 1975; Melrose, 1980;
Webb, 1983), i.e., 
\beq
  D_p \simeq \frac{p^2}{3 \tau} \left(\frac{V_A}{c}\right)^2\,,
\eeq where $V_A =B/\sqrt{4\pi\rho}$ is the Alfven velocity. This 
implies a characteristic acceleration timescale
\beq\label{acc}
t_{\rm acc}=p^3 \left[\frac{\pd}{\pd p}\left(p^2 D_p\right)
            \right]^{-1}
           = \frac{3}{(4-\alpha)}\left(\frac{c}{V_A}\right)^2 \tau
\eeq for $\tau = \lambda/c \propto p^{\alpha}$. Comparison with 
the results for diffusive shock acceleration shows that second-order 
Fermi acceleration is typically a factor of order $(u_s/V_A)^2$ 
slower than first-order Fermi acceleration.\\
Second-order Fermi acceleration is usually expected to lead to 
particle spectra $N(\gamma)\propto \gamma^{-s}$ that are typically 
flatter than those produced by standard strong shock acceleration, 
i.e., $s<2$ (cf. also Virtanen and Vainio, 2005). For example, 
adding both a monoenergetic source term $Q\,\delta(p-p_0)$ and a 
particle loss term $-f/T$ on the rhs of eq.~(\ref{diffusion}), 
assuming $\alpha=0$ and steady state, one finds 
\beq
  s=\frac{3}{2} \sqrt{1+16\,t_{\rm acc}/(9\,T)} -\frac{1}{2}
\eeq above $p_0$, so that for $T \gg t_{\rm acc}$ for example, one 
has $N(\gamma)\propto \gamma^{-1}$.

\subsection{Gradual shear acceleration}
Consider now the case where the magnetic turbulence structures
are embedded in a gradual shear flow $\vec{u}=u_z(x)\vec{e}_z$,
assuming their random velocities to be small compared to the
characteristic shear velocity. Particles traveling across the
shear thus encounter scattering centres with different (although
non-random) local velocities $u_z(x)$. Similar to the case of 2nd
order Fermi acceleration, the average energy gain per collision
becomes second order in $\tilde{u}/c$, where $\tilde{u} = (\pd u_z/
\pd x) \lambda$ denotes the characteristic relative velocity of
the scattering centres, $\lambda \simeq c \tau$ is the particle
mean free path and $\tau$ is the mean scattering time, i.e., one
finds (cf. also Jokipii and Morfill, 1990)
\beq
 \frac{<\Delta \epsilon>}{\epsilon_1} \propto 
               \left(\frac{\pd u_z}{\pd x}\right)^2 \tau^2\,.
\eeq
Again, as a stochastic process shear acceleration can be described
by a diffusion equation in momentum space, i.e., an equation of
type eq.~(\ref{diffusion}). For nonrelativistic gradual shear 
flows a proper statistical treatment (Rieger and Duffy, 2006) 
gives 
\beq 
D_p = \tilde{\Gamma} p^{2} \tau \propto p^2 \tau\,,
\eeq where $\tilde{\Gamma}=(\pd u_z/\pd x)^2/15$ is the shear 
flow coefficient and $\tau =\tau_0\,p^{\alpha}$ is the mean 
scattering time. This implies a characteristic acceleration 
timescale [cf. eq.~(\ref{acc})]
\beq\label{shear}
t_{\rm acc} = 1/([4+\alpha]\,\tilde{\Gamma}\,\tau)
\eeq which, in contrast to first- and second-order Fermi, is
inversely proportional to the particle mean free path $\lambda
\simeq \tau/c$. Eq.~(\ref{shear}) can be generalized to the
relativistic case by replacing $\tilde{\Gamma}$ by its 
relativistic counterpart (Rieger and Duffy, 2004, eq.(3)). 
In particular, for a (cylindrically collimated) shear flow 
decreasing linearly with radial coordinate from relativistic 
to nonrelativistic speeds over a distance $\Delta r=(r_2-r_1)$, 
the maximum acceleration timescale is of order 
\beq\label{t_shear}
t_{\rm acc} \simeq \frac{3 (\Delta r)^2}{\gamma_b(r_1)^4 
           \lambda\,c}\,\
\eeq where $\gamma_b(r)$ is the local bulk Lorentz factor 
of the flow.\\
In the simplest case, assuming quasi steady state conditions 
and monoenergetic injection, the functional form of the local 
particle distribution $N(p) \propto p^2\,f(p)$ becomes 
\beq
 N(p) \propto p^{-(1+\alpha)}
\eeq above $p_0$ for $\alpha >0$ (Berezhko and Krymskii, 
1981; Rieger and Duffy, 2006). From an astrophysical point of 
view, acceleration becomes essentially non-gradual when the 
particle mean free path becomes larger than the width of the 
transition layer. In this case results from the study of 
relativistic non-gradual shear flows (e.g., Ostrowski, 1990) 
can be employed to analyse issues of maximum energies and 
resulting particle distributions.

\section{Application to astrophysical jet sources}
Apart from lateral particle escape and limited jet activity, radiative
synchrotron losses represent one of the most servere constraints for 
the acceleration of energetic particle in astrophysical jets. We may
estimate the maximum achievable particle energy in the presence of 
synchrotron losses by equating the isotropic synchrotron cooling 
timescale $t_{\rm cool}=(9\,m^3\,c^5)/(4\,\gamma\,e^4\,B^2)$ with 
the corresponding acceleration timescale. For non-relativistic shock 
acceleration this results in a maximum comoving Lorentz factor 
(provided $r_g(\gamma_{\rm max})$ is still smaller than the width 
of the jet) of order
\beq\label{g_shock}
   \gamma_{\rm max}^{1st} \simeq 9\cdot 10^9 \left(\frac{1\, 
                     \mathrm{G}}{B}\right)^{1/2}
                     \left(\frac{m}{m_p}\right)\,
                     \left(\frac{u_s}{0.1\,\mathrm{c}}\right)\,
\eeq with $m_p$ the proton mass, whereas in the case of second order 
Fermi with $\lambda \simeq r_g$ one finds
\beq\label{g_2nd}
   \gamma_{\rm max}^{2nd} \simeq 2\cdot 10^8 \left(\frac{1\, 
                          \mathrm{G}}{B}\right)^{1/2}
                          \left(\frac{m}{m_p}\right)\,
                          \left(\frac{V_A}{0.001\,\mathrm{c}}\right)\,\,.
\eeq For $\lambda \sim r_g$ the acceleration timescale in a gradual 
shear flow [e.g., eq.~(\ref{t_shear})] scales with $\gamma$ in the 
same way as the cooling timescale, so that radiative losses are no 
longer able to stop the acceleration process once it has started to 
operate efficiently. For a linearly decreasing flow profile efficient
acceleration thus becomes possible if the shear is sufficiently strong,
i.e., provided the relation
\beq\label{r_shear}
  \Delta r \lppr 0.1 \,\gamma_b(r_1)^2 \left(\frac{m}{m_p}\right)^2 
           \left(\frac{1\, \mathrm{G}}{B}\right)^{3/2}\,\,\mathrm{pc}\,
\eeq holds. These considerations suggest the following:

\subsection{Relativistic AGN jets}
In the case of relativistic AGN jets, diffusive shock acceleration 
({\it first order Fermi}) processes represent the most efficient and 
plausible mechanism for the origin of the observationally required 
non-thermal powerlaw distributions in their inner (sub-parsec scale) 
jets. For characteristic parameters, e.g., $u_s \sim 0.1$ c, $B \sim 
0.1\, b_0$ Gauss, maximum electron Lorentz factors $\gamma_{\rm max} 
\lppr 10^7\,b_0^{-1/2}$ [cf. eq.~(\ref{g_shock})] may be reached 
suggesting that for blazar-type sources with $\gamma_b \sim 10$ an 
electron synchrotron contribution up to $\nu_{\rm obs} \sim 10^5 
\gamma_b\,\gamma_{\rm max}^2\,b_0 < 2 \cdot 10^{20}\,$ Hz, i.e., well 
in the hard X-ray regime, may be possible, provided IC losses do not 
dominate. Furthermore, the variability in the high energy regime may 
be very fast if associated either with $t_{\rm acc}$ or $t_{\rm 
cool}$. One expects a similar expression for the acceleration 
timescale [eq.~(\ref{t_shock})] to hold if shocks in blazar-type 
jets are mildly relativistic ($u_s \gppr 0.3$ c), although the 
spectral index may then be somewhat steeper, i.e, $s > 2$, and the 
explicit results more dependent on the exact scattering conditions 
(e.g., Lemoine and Pelletier, 2003).\\
As noted in the introduction (e.g., see the case of the famous 
quasar 3C~273), the situation is somewhat different with respect 
to the (collimated) relativistic large-scale jets in AGNs,
\footnote{There is now mounting observational evidence that the 
jets of powerful FR~II type sources are still relativistic ($\gamma_b 
\sim 5$) on large kpc scales (e.g., Sambruna et al., 2001; 
Tavecchio et al., 2000).} where observational evidence suggests that 
shock acceleration is not sufficient to account for the observed smooth 
evolution of the spectral index. Stochastic acceleration like shear 
or second-order Fermi processes may represent the most natural 
candidates for distributed acceleration mechanisms operating all 
along the jet (Stawarz and Ostrowski, 2000; Rieger and Duffy, 2004). 
For a quasi-uniform flow profile with $\gamma_b \sim (3-5)$ and 
a characteristic set of parameters, $B \sim 10^{-5} b_0$ G, $b_0 
\gppr 1$ (e.g., Stawarz, 2005) and $V_A \sim 10^8\, b_0$ cm/s, 
{\it second-order Fermi} acceleration [eq.~(\ref{g_2nd})] may account 
for electrons with maximum Lorentz factor up to $\gamma_{\rm max} 
\sim 10^8\,b_0^{1/2}$, corresponding to synchrotron emission up 
to $\nu_{\rm obs} \sim 5 \cdot 10^{17}\, b_0^2 \,\mathrm{Hz} 
\lppr 2\,b_0^2$ keV. Neglecting inverse Compton losses for a 
moment, it seems possible that synchrotron emission from 
electrons accelerated via second-order Fermi processes can, at 
least in principle, account for the observed extended emission in 
the optical (Jester et al., 2001), and perhaps even in the 
Chandra X-ray regime (cf. Harris and Krawczynski, 2006). If this 
is indeed the case, the radiating electron distribution $N(\gamma) 
\propto \gamma^{-s}$ is likely to consist of at least two components: 
One, localized at the observed knots and corresponding to strong 
shock acceleration with spectral index $s \simeq 2$, and one, 
distributed in between knots and associated with second order 
Fermi processes and flatter spectral index $s<2$. It is likely, 
however, that the real situation is much more complex. There is 
strong evidence, for example, that in reality astrophysical 
jets do not possess a simple uniform flow profile as often used 
for spectral modelling. In particular, the density and velocity 
gradients associated with extreme astrophysical environments 
are likely to result in a non-negligible velocity shear across 
the jet. In the case of AGN such shear flows are indeed
observationally well-established, e.g., see Laing \& 
Bridle~(2002) and Laing et al.~(2006) for recent observational 
results and modelling, and Rieger \& Duffy (2004) for a recent 
review of the phenomenological evidence. For the characteristic 
parameters specified above, eq.~(\ref{r_shear}) then suggests that 
on kiloparsec scales efficient {\it shear acceleration} of 
electrons may be possible provided the velocity decreases 
significantly on radial scales $\Delta r_e \lppr$ several 
percent of the total jet width $r_j \sim 1$ kpc. Essentially no 
such constraint applies to protons (i.e., $\Delta r_p \sim r_j$,
cf. eq.~[\ref{r_shear}]), so that compared to the case of 
electrons efficient proton acceleration should be much more 
common. These considerations have interesting observational 
consequences: (i) Suppose that the velocity shear in the 
large-scale jet is sufficiently strong (e.g., significant 
velocity decay over $\Delta r_e \sim$ several percent of 
$r_j$) to allow for efficient electron acceleration, 
the required high energy seed particles ($\gamma \gppr 10^6\,b_0$) 
being provided by first and second-order Fermi processes. Even for 
the simplistic case of $V_A$ independent of $r$, shear acceleration 
will then begin to dominate over second-order Fermi processes for 
electrons with $\gamma \gppr 10^8\,(\Delta r/10\, \mathrm{pc})\,
(V_A/0.01\,c) \,b_0$, resulting in a third radiating electron 
component with local index $s \simeq 2$. If this is indeed the case, 
the observed spectral index in the optical-UV (probably due to 
second-order Fermi acceleration) may well be different from the 
one measured in the X-ray regime (likely due to shear acceleration). 
(ii) On the other hand, even if the large-scale shear is very weak 
(e.g., say $\Delta r \sim 0.3$ kpc) efficient acceleration of 
protons remains still possible suggesting that relativistic 
gradual shear flows may allow acceleration of protons up to 
ultra-high energies of $10^{18}-10^{19}$ eV (where $\lambda
\sim \Delta r$), i.e., well up to the ``angle'' of the cosmic 
ray energy spectrum at around  $3 \times 10^{18}$ eV. 
Subsequent non-gradual shear acceleration may then reach even 
higher proton energies, perhaps even up to $\sim 10^{20}$ eV 
(Ostrowski, 1990 and 1998). If so, then one would naturally 
expect a change in spectral index around the angle when gradual 
shear is replaced by non-gradual shear acceleration usually 
associated with flatter particle spectra.

\subsection{Ultrarelativistic GRB outflows}
\begin{figure}
\centering
  \includegraphics[width=0.49\textwidth]{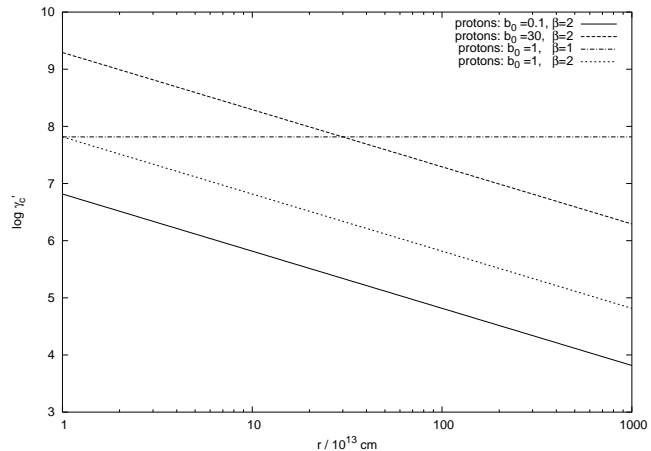}
\caption{Plot of the critical (comoving) Lorentz factor 
$\gamma_c'$ as a function of distance $r$ assuming an internal 
shock speed of $\beta_s=0.95$ c. Above $\gamma_c'$ the shear 
acceleration timescale becomes smaller than the shock 
acceleration timescale, so that proton acceleration by shear 
becomes more efficient.}
\label{plot1}       
\end{figure}

While there is strong evidence today that GRBs are associated 
with collimated ultra-relativistic outflows or jets (Rhoads, 
1999; Kulkarni et al., 1999; Greiner et al., 2003), it is still 
a matter of ongoing debate whether these jets exhibit a rather
uniform ("top-hat") or a more universal structured ("power-law" 
or "Gaussian"-type) hydrodynamical profile (e.g., Rhoads, 1999; 
Rossi et al., 2002). In any case, it is commonly believed that 
shock accelerated electrons are, via synchrotron radiation 
processes, responsible for both, the prompt gamma-ray burst and 
its afterglow emission (Piran, 2005): Efficient electron 
acceleration at mildly relativistic internal shocks ($\Gamma_s$ 
of a few) arising from velocity variations in the relativistic 
outflow, is usually thought to be behind the powerful burst of 
$\gamma$-rays (Rees and M{\' e}sz{\' a}ros, 1994), while 
electron acceleration at a decelerating, highly relativistic 
($\Gamma_s \gg 1$) external shock is believed to be responsible 
for the afterglow emission, peaking successively in the 
$\gamma$-rays, X-rays, optical and the radio regime (Rees and 
M{\' e}sz{\' a}ros, 1992). Whereas to order of magnitude 
accuracy, the acceleration timescale at a mildly relativistic 
shock front may be reasonably approximated by the Larmor time, 
i.e., $t_{\rm acc} \sim r_g/c$, the acceleration timescale at 
highly relativistic shocks may be as short as a fraction $1/
\Gamma_s$ of the (upstream) Larmor time (e.g., Gallant and 
Achterberg, 1999; Lemoine and Pelletier, 2003). 
The latter result is related to the fact that (for all but 
the initial crossing) particles upstream cannot be deflected 
beyond an angle $\sim 1/\Gamma_s$ before being overtaken by 
the shock. Accordingly, particles can also only gain a factor 
of order $\Gamma_s^2$ in energy in the first shock crossing 
cycle, whereas the energy gain is reduced to a factor of order 
$2$ for subsequent crossing events as particles upstream do not 
have sufficient time to isotropise (Gallant and Achterberg, 
1999; Achterberg et al., 2001; cf. however also Derishev et 
al., 2003). In general the power-law spectral index of the 
accelerated particle distribution for ultra-relativistic 
shock acceleration is strongly dependent on the exact 
scattering conditions. In the case of pitch-angle 
diffusion the simulations give $s \sim 2.2 - 2.3$ (e.g., 
Bednarz and Ostowski, 1998; Kirk and Duffy, 1999; Achterberg 
et al., 2001; Baring, 2005).\\
It has been suggested by Waxman, that the first- (Waxman, 2004) 
and perhaps also second-order (Waxman, 1995) Fermi processes, 
known to accelerate electrons in GRB outflows with $\gamma_b
\sim 300$ to gamma-radiating energies, may also allow an 
efficient acceleration of protons to ultra-high cosmic ray 
energies in excess of $10^{20}$ eV. 
While this seems in principle possible, the analysis of the 
natural shear acceleration potential in expanding 
ultra-relativistic GRB flows (Rieger and Duffy, 2005) suggests 
that under a reasonable range of conditions shear acceleration 
may be more relevant for the production of UHE cosmic rays 
than shock-type acceleration processes. This is illustrated in 
Fig.~\ref{plot1}, where we have plotted the critical (comoving) 
proton Lorentz factor $\gamma_c'$, defined by $t_{\rm acc}(\rm 
shear) = t_{\rm acc}(\rm shock)$, as a function of distance $r$, 
assuming an intrinsic magnetic field strength $B=1000\,b_0$ $
(10^{13}\, \mathrm{cm}/r)^{\beta}$ and $\eta\,\xi \simeq 1$ 
(cf. Rieger and Duffy, 2005, for more details), noting that in 
order to achieve UHE greater than $10^{20}$ eV, characteristic 
(comoving) proton Lorentz factors of $\gppr 4 \cdot 10^8$ are
required 

\subsection{Mildly relativistic Microquasar jets}
Similar to the case of extragalactic AGN jets, Fermi-type processes 
are likely to lead to efficient particle acceleration in the scale
down versions of AGNs known as galactic Microquasars (MQs) (Mirabel 
and Rodr{\'{\i}}guez, 1999). MQs are (radio-loud) X-ray binary systems 
where a compact object (a neutron star or stellar mass black hole) 
accretes matter from a normal star in orbital motion around it and 
(at least in the low-hard state) produces a (quasi-stable) collimated, 
mildly relativistic jet with characteristic bulk velocities in the 
range of $0.2-0.9$ c (cf. Gallo et al., 2003) and observed jet lengths 
in some sources well in excess of several hundred AUs. The detection of 
extended, nonthermal radio emission in MQs substantiate the presence 
of relativistic electrons in their jets. Indeed, detailed modelling 
reveals that synchrotron emission from relativistic electrons is 
likely to be important in the radio up to the soft gamma-ray regime 
for a magnetic field close to equipartition (Bosch-Ramon et al., 2006). 
Assuming a magnetic field scaling $B(z) = 10^5\,b_0\,(z_0/z)$ G, with 
$b_0 \sim 1$ and $z_0 \sim 50\,R_g$ (e.g., Bosch-Ramon et al., 2006), 
synchrotron-limited, non-relativistic shock processes ($u_s \sim 0.1\,$c) 
give electron Lorentz factors $\gamma_e \lppr 1.5 \cdot 10^4\, 
b_0^{-1/2}\,(z/z_0)^{1/2}$, cf. eq~(\ref{g_shock}), thus allowing for 
electron synchrotron emission up to $\nu_{\rm obs} \sim 2 \cdot 
10^{19}$ Hz ($\sim 10^5$ eV), whereas the condition of lateral 
confinement limits possible maximum Lorentz factors for protons 
(electrons) to $\gamma_p \lppr 7 \cdot 10^4\,b_0$ ($\gamma_e \lppr 
10^8\,b_0$), assuming a typical half-opening angle $\phi \simeq 0.05$ 
rad. Shock acceleration alone seems thus not to be able to provide the 
required high-energy electrons suggested by spectral modelling results, 
e.g., see the required high acceleration efficiency in the case of 
LS 5039 (Parades et al.~2006). Interestingly, 
however, post-acceleration of electrons, occurring on scales larger 
than $z/z_0 \sim 10$ [cf. eq.~(\ref{r_shear})] within a strong 
($\Delta r_e \sim$ several percent of $r_j$) velocity shear 
may provide a possible solution by boosting shock-accelerated 
electrons further up to maximum Lorentz factors $\gamma_e \sim 10^6$, 
substantiating the notion that SSC (Klein-Nishina) or star IC (Thomson) 
may yield VHE $\gamma$-rays possibly reaching several TeV energies 
(Paredes et al., 2000). We note that further evidence for particle
acceleration beyond standard first-order Fermi has been suggested 
recently (Gupta and B\"ottcher, 2006).

\section{Conclusions}
While the relations derived above essentially rely on a simple 
test particle approach, thereby neglecting any back-reaction 
effects of the accelerated particles (e.g., strong shock 
modification, viscous kinetic energy dissipation or significant 
wave damping), we believe that they still allow reasonable order 
of magnitude estimates for many cases of interest. Hence, although 
there are different physical conditions in the relativistic jets 
of AGNs, Microquasars and GRBs, our analysis suggests that 
Fermi acceleration processes offer a powerful and attractive 
explanatory framework for the origin of the non-thermal particle 
distributions required within these sources. 
In particular, due to its inverse scaling, $t_{\rm acc} \propto 
1/\lambda$, shear acceleration is likely to become important at 
high energies and may thus naturally lead to the presence of an 
at least two-component energetic particle distribution.

\begin{acknowledgements}
Support by a Cosmogrid Fellowship (FMR) is gratefully acknowledged.
\end{acknowledgements}

\end{document}